\documentclass[twocolumn,pra,showpacs,aps,superscriptaddress]{revtex4}

\usepackage{amsmath}
\usepackage{amssymb}
\usepackage{latexsym}
\usepackage{graphicx}
\usepackage[dvips]{color}
\begin{document}
\title{Scaling of noise correlations in one-dimensional-lattice-hard-core-boson systems}

\author{Kai He}
\affiliation{Department of Physics, Georgetown University, Washington, DC 20057, USA}
\author{Marcos Rigol}
\affiliation{Department of Physics, Georgetown University, Washington, DC 20057, USA}
\affiliation{Kavli Institute for Theoretical Physics, University of California, Santa Barbara,
California 93106, USA}

\begin{abstract}
Noise correlations are studied for systems of hard-core bosons in
one-dimensional lattices. We use an exact numerical approach based
on the Bose-Fermi mapping and properties of Slater determinants. We
focus on the scaling of the noise correlations with system size in
superfluid and insulating phases, which are generated in the
homogeneous lattice, with period-two superlattices and with
uniformly distributed random diagonal disorder. For the superfluid
phases, the leading contribution is shown to exhibit a density-
independent scaling proportional to the system size, while the first
subleading term exhibits a density-dependent power-law exponent.
\end{abstract}
\maketitle

\section{Introduction}
In the past decade, ultracold quantum gases have gained considerable attention due to the unique control
achieved experimentally for manipulating such systems. This has enabled experimentalists to explore the
richness and complexity of strong correlations and reduced dimensionalities and to even simulate model
Hamiltonians used to understand complicated materials \cite{bloch_dalibard_review_08}. For the goal of
studying quantum systems in quasi-one-dimensional geometries, remarkable experimental examples include
the realization of quantum gases in very anisotropic traps \cite{schreck_khaykovich_01,gorlitz_vogels_01}
and loading Bose-Einstein condensates in deep two-dimensional optical lattices
\cite{greiner_bloch_01,moritz_stoferle_03,stoferle_moritz_04,paredes_widera_04,kinoshita_wenger_04,%
kinoshita_wenger_05} and in atom chips \cite{trebbia_esteve_06,hofferberth_lesanovsky_07,amerongen_es_08}.

One dimension hosts a variety of models that can be exactly solved analytically and as such
are of very much interest to both theorists and experimentalists. Remarkably, the high degree of tunability
and isolation achieved in ultracold gases experiments has permitted the realization of various such models.
An example of particular relevance to the work presented here was the realization of a gas of
impenetrable bosons (hard-core bosons), also called a Tonks-Girardeau gas, in the presence \cite{paredes_widera_04}
and absence \cite{kinoshita_wenger_04,kinoshita_wenger_05} of a lattice along the one-dimensional gas.
The problem of indistinguishable impenetrable bosons in one dimension was first analyzed by Girardeau
\cite{girardeau_60}, who noticed that its thermodynamic properties could be easily computed by mapping
such a problem to that of indistinguishable noninteracting spinless fermions. In the presence of a lattice,
the hard-core boson problem (see below) can be mapped to a special case of the $XY$ spin-1/2 chain introduced
by Lieb {\it et al}. \cite{lieb_61}, whose thermodynamic and local properties can also be solved by
mapping it to a noninteracting spinless fermion lattice model.

The calculation of the off-diagonal correlations, such as the one-particle correlations, is a more
challenging task. In the homogeneous case, this has been done in various works and using
various approaches for both continuous and lattice systems
\cite{lenard_64,mccoy_68,vaidya_tracy_78,vaidya_tracy_79a,jimbo_80,gangardt_04}. It should be noted, however,
that the experimental realization of these model Hamiltonians requires a trap for containing the gas. This means
that such experimental systems are in general inhomogeneous and their description requires one to take into
account the presence of the trapping potential, which is to a good approximation harmonic. Studies of
one-particle correlations of harmonically trapped Tonks-Girardeau gases have been performed in a series of
more recent works
\cite{girardeau_wright_01,lapeyre_girardeau_02,papenbrock_03,forrester_03b,gangardt_04,rigol_muramatsu_04HCBa,%
rigol_muramatsu_05HCBb,rigol_05}.

One-particle correlations can be probed in experiments by means of time-of-flight measurements, in which
the confining potentials are turned off and, in the absence of interactions during the expansion, the initial
momentum distribution of the trapped gas is mapped onto the density distribution of the system after a long
expansion time. The latter density distribution is then measured by taking a picture of the gas after expansion.
How the scaling of the one-particle correlations in the trapped system is reflected in the momentum distribution,
which is the diagonal part of the Fourier transform of the one-particle density matrix, was also discussed in
several works mentioned above \cite{girardeau_wright_01,lapeyre_girardeau_02,papenbrock_03,rigol_muramatsu_04HCBa,%
rigol_muramatsu_05HCBb,rigol_05}.

Remarkably, it was also proposed that higher order correlations can be measured after time of flight
by analyzing the atomic shot noise in the images \cite{altman_demler_04}. These noise correlations are
experimentally associated with Hanbury-Brown-Twiss interferometry, which allow one to measure the
density-density correlations in the spatial images. After long expansion times, under the usual assumption
of absence of interactions during the expansion, noise correlations reflect the momentum space
density-density correlations in the trapped system. Shortly after the theoretical proposal \cite{altman_demler_04},
noise correlations were measured in experiments with bosons in three-dimensional optical lattices
\cite{folling_gerbier_05} and with attractive fermions \cite{greiner_regal_05}.

Our goal in this paper is to explore the scaling of the noise correlations in various ground-state
phases of one-dimensional hard-core-boson-lattice systems. We consider the homogeneous case, systems
with an additional period-two superlattice potential, and disordered systems with a uniform random
distribution of local potentials. We implement an exact numerical approach to compute the noise correlations,
which follows after the hard-core-boson-lattice model is mapped onto a noninteracting spinless fermion
model by means of the Holstein-Primakoff transformation \cite{holstein_primakoff_40} and the Jordan-Wigner
transformation \cite{jordan_28}. This approach is an extension of the method developed by one of the authors
(in collaboration with A. Muramatsu) \cite{rigol_muramatsu_04HCBa,rigol_muramatsu_05HCBb} for the exact
calculation of the one-particle density matrix of hard-core-boson-lattice systems using properties of Slater
determinants. We should note that earlier studies of noise correlations in hard-core-boson-lattice models followed
an alternative numerical formulation based on Wick's theorem \cite{rey_satija_06a,rey_satija_06b,rey_satija_06c}.
However, the lattice sizes accessible within that approach were too small to enable a systematic study of the
scaling of the noise correlations with system size.

There are three ground-state phases on which we will focus our present study, which are the
superfluid phase, the Mott-insulating or charge-density-wave phase, and the Anderson-glass phase. Those phases
can be obtained in the various background potentials mentioned before. In the superfluid phase,
we show that the leading contribution
to the noise correlation peaks scales linearly with the size of the system, independently of the density and of
the absence or presence of a superlattice potential, while the first subleading term does depend on both.
As expected, for the Mott and Anderson-glass phases, which are both insulating, the scaling of the peaks shows an
asymptotic value that depends on the density and strength of the background potential but that is independent
of the system size. The leading-order results are consistent with the behavior of the zero-momentum peak of the
momentum distribution, which scales with the square root of the system size in the superfluid phases
\cite{lenard_64} while it saturates in the insulating \cite{rousseau_arovas_06,rigol_muramatsu_06} and disordered
\cite{horstmann_cirac_07} phases. The latter behavior is a result of the short-range correlations present
in the insulating phases.

This presentation is organized as follows. In Sec.~\ref{sec:ExactApproach}, we describe the models and introduce
the exact numerical approach. In Secs.~\ref{sec:periodic}, \ref{sec:period-2}, and \ref{sec:disorder}, we study
the noise correlations in the homogeneous case, in period-two superlattices, and in disordered systems, respectively.
A comparison between the noise correlations in all those systems is also presented in Sec.\ \ref{sec:disorder}.
Finally, Sec.~\ref{sec:Conclusions} summarizes our results.

\section{Exact Approach}\label{sec:ExactApproach}
\subsection{Hamiltonian and relevant quantities}

In the hard-core limit of the Bose-Hubbard model, the one-dimensional Hamiltonian can be written as
\begin{equation}
\label{HamHCB} \hat{H}_\textrm{HCB} = -t \sum_{i} \left(\hat{b}^\dagger_{i} \hat{b}^{}_{i+1}
+ \textrm{H.c.} \right) + \sum_{i} V_{i} \hat{n}_{i},
\end{equation}
where $t$ represents the hopping parameter and $\{V_{i}\}$ a set of on-site potentials. The hard-core boson
creation and annihilation operators at site $i$ are denoted by $\hat{b}^\dagger_{i}$ and $\hat{b}^{}_{i}$,
respectively, and $\hat{n}_{i}=\hat{b}^\dagger_{i} \hat{b}^{}_{i}$ denotes the occupation operator of site $i$.
While the bosonic commutation relations $[\hat{b}^{}_{i}, \hat{b}^\dagger_{j}]=\delta_{ij}$ still hold for all
sites, additional on-site constraints apply to the creation and annihilation operators
\begin{equation}
\label{ComHCB} \hat{b}^{\dagger 2}_{i}= \hat{b}^2_{i}=0,
\end{equation}
which preclude multiple occupancy of the lattice sites. Note that Eq.\ (\ref{ComHCB}) is only valid when
applied to a string of bosonic operators in normal order \cite{rey_satija_06a}, as will be explained below.

The hard-core-boson Hamiltonian can be mapped onto the exactly solvable noninteracting fermion Hamiltonian
by means of Bose-Fermi mapping, which follows in two steps. The first step is given by the correspondence
between hard-core bosons and spin-1/2 systems through the Holstein-Primakoff transformation
\cite{holstein_primakoff_40}
\begin{eqnarray}
\hat{\sigma}^+_{i}& =& \hat{b}^\dagger_{i}\; \sqrt{1-\hat{b}^\dagger_{i} \hat{b}^{}_{i}}, \quad
\hat{\sigma}^-_{i} = \sqrt{1-\hat{b}^\dagger_{i} \hat{b}^{}_{i}}\; \hat{b}^{}_{i},\nonumber\\
\hat{\sigma}^z_i&=& \hat{b}^\dagger_{i} \hat{b}^{}_{i} -\frac{1}{2}, \label{HPT}
\end{eqnarray}
where $\hat{\sigma}^\pm_{i}$ are the spin raising and lowering operators and $\hat{\sigma}^z_{i}$
is $z$-component Pauli matrix for spin-1/2 systems. A straighforward analysis reveals that
$\hat{b}^\dagger_i (\hat{b}^{}_i)$ can be directly replaced by $\hat{\sigma}^+_i (\hat{\sigma}^-_{i})$
if and only if the hard-core boson creation and annihilation operators are arranged in normal order; that is,
all creation operators must be placed to the left of the annihilation operators before the mapping.
The root of this difference between hard-core-boson and spin-1/2 systems lies in the fact that despite
 the suppressed multiply-occupied states, virtual states of multiple occupancy can occur in the infinite
$U$ limit of the Bose-Hubbard model and they need to be properly taken into account for a correct
calculation of bosonic correlations \cite{rey_satija_06a}. As mentioned above, in general Eq.\ (\ref{ComHCB})
does not apply, for example, for a bosonic system (independently of the value of $U$):
$\langle 0|b\, b \,b^\dag b^\dag|0 \rangle = \langle 1|b \: b^\dag|1 \rangle = 2$, and a direct
replacement of the hard-core-boson operators by spin operators would lead to a strictly zero expectation value.
To avoid this problem, the proper recipe is to normal order strings of hard-core-boson operators using the bosonic
commutation relations before making the replacement
$\hat{b}^\dagger_i (\hat{b}^{}_i)\rightarrow\hat{\sigma}^+_i (\hat{\sigma}^-_{i})$.

In the second step, the spin-1/2 Hamiltonian can be mapped onto a noninteracting
fermion Hamiltonian by means of Jordan-Wigner transformation \cite{jordan_28},
\begin{eqnarray}
 \label{JWT} \hat{\sigma}^{+}_i&=&\hat{f}^{\dag}_i
\prod^{i-1}_{\beta=1}e^{-i\pi \hat{f}^{\dag}_{\beta}\hat{f}^{}_{\beta}},\quad
\hat{\sigma}^-_i=\prod^{i-1}_{\beta=1} e^{i\pi \hat{f}^{\dag}_{\beta}\hat{f}^{}_{\beta}}\hat{f}_i,\nonumber\\
\hat{\sigma}^z_i&=& \hat{f}^\dagger_{i} \hat{f}^{}_{i} -\frac{1}{2},
\end{eqnarray}
with $\hat{f}^\dag_i$ and $\hat{f}^{}_i$ being the creation and annihilation operators for spinless fermions,
respectively.

The noninteracting fermions share the exact same form of the Hamiltonian as the hard-core bosons
up to a boundary term that for periodic systems depends on whether the total number of bosons $N$ in the system
is even or odd \footnote{For periodic hard-core boson chains, the equivalent noninteracting fermion Hamiltonian
satisfies periodic boundary conditions if the total number of hard-core bosons is odd and antiperiodic boundary
conditions if the total number of hard-core bosons is even \cite{lieb_61}.}:
\begin{equation}
\label{HamFerm} \hat{H}_\textrm{F} =-t \sum_{i} \left( \hat{f}^\dagger_{i}
\hat{f}^{}_{i+1} + \textrm{H.c.} \right)+ \sum_{i} V_i \,\hat{n}^f_{i},
\end{equation}
where $\hat{n}^f_i=f^\dag_{i} f^{}_i$ is the fermionic occupation operator of site $i$. This mapping
shows that all thermodynamic properties and real space density-density correlations of hard-core bosons
are identical to those of a system of noninteracting fermions. This is of course not true for the off-diagonal
correlation functions.

In order to compute the one-particle correlations, one can follow the approach described in
Refs.\ \cite{rigol_muramatsu_04HCBa,rigol_muramatsu_05HCBb}. (Note that in those studies the hard-core boson
and spin-1/2 operators were used indistinctively but consistently with the discussion here.) One can write
$\hat{\rho}_{ij} = \hat{b}^\dagger_i\hat{b}_j=\hat{\sigma}^+_i\hat{\sigma}^-_j$ and
\begin{equation}\label{SpinComm}
\hat{\sigma}^+_i\hat{\sigma}^-_j=\delta_{ij}+(-1)^{\delta_{ij}} \hat{\sigma}^-_j\hat{\sigma}^+_i,
\end{equation}
so that to compute the one-particle density matrix
$\rho_{ij}=\langle\hat{\rho}_{ij}\rangle$ one only needs to
calculate
\begin{eqnarray}
\label{eq:greensHCB}
G_{ij}&=&\langle \hat{\sigma}^{-}_i\hat{\sigma}^+_j\rangle=\langle\Psi_{F}|\prod^{i-1}_{\beta=1}
e^{i\pi \hat{f}^{\dag}_{\beta}\hat{f}^{}_{\beta}} \hat{f}^{}_i \hat{f}^{\dag}_j
\prod^{j-1}_{\gamma=1} e^{-i\pi \hat{f}^{\dag}_{\gamma}\hat{f}^{}_{\gamma}}
|\Psi_{F}\rangle\nonumber\\
&=&\det\left[ \left( {\bf P}^{i} \right)^{\dag}{\bf P}^{j}\right],
\end{eqnarray}
where
\begin{equation}\label{eq:sd}
|\Psi_{F}\rangle=\prod^{N}_{\kappa=1} \sum^L_{\varrho=1} P_{\varrho \kappa}\hat{f}^{\dag}_{\varrho}\ |0 \rangle
\end{equation}
is the Slater determinant corresponding to the fermionic wave-function ($L$ is the number of lattice sites),
and $({\bf P}^{\alpha})_{L,N+1}$, with $\alpha=i,j$, is obtained using properties of Slater determinants and written as
\begin{eqnarray}\label{eq:matcomp}
 P^{\alpha}_{\varrho \kappa}= \left\{ \begin{array}{rl}
 -P_{\varrho \kappa} & \text{for } \varrho<    \alpha,\,\kappa=1,\ldots,N \\
\,P_{\varrho \kappa} & \text{for } \varrho\geq \alpha,\,\kappa=1,\ldots,N \\
  \delta_{\alpha\varrho} & \text{for } \kappa=N + 1
\end{array}\right.
\end{eqnarray}

Once $\rho_{ij}$ is computed, the momentum distribution function can be determined using the Fourier transform
\begin{equation}
\label{MomDist} n_k=\frac{1}{L} \sum_{i j} e^{i ka(i-j)} \rho_{i j},
\end{equation}
where $a$ is the lattice constant.

\subsection{Noise correlations}

In this work we are interested in the second-order correlations of hard-core boson systems in the
quasi-momentum space \cite{altman_demler_04}. These noise correlations are defined as
\begin{equation}
\label{NoiseDef} \Delta_{k k^\prime} \equiv \langle \hat{n}_k \hat{n}_{k^\prime} \rangle \
- \langle \hat{n}_k\rangle \langle \hat{n}_{k^\prime}\rangle - \langle \hat{n}_k\rangle \
\delta_{k-k^\prime, nK},
\end{equation}
where $K=2\pi/a$ is the reciprocal lattice vector and $n$ is a nonzero integer. The second and third terms
in Eq.\ (\ref{NoiseDef}) can be computed using the approach mentioned in the previous subsection,
so we focus here on how to compute the first term
\begin{equation}
\label{NoiseCorr} \langle \hat{n}_k \hat{n}_{k^\prime} \rangle = \frac{1}{L^2} \sum_{i j l m}\\
e^{i k a (i-j)+i k^\prime a (l-m)} \langle \hat{b}^\dag_i \hat{b}_j
\hat{b}^\dag_l \hat{b}_m \rangle,
\end{equation}
for which we extend the recipe for calculating the two-point correlations
\cite{rigol_muramatsu_04HCBa,rigol_muramatsu_05HCBb} to obtain four-point correlations and hence
the noise correlations.

From the mapping between hard-core bosons and spins one gets the following expression for the four-point
correlation function in terms of spin operators
\begin{eqnarray}
\label{4pCorrHCB}
\langle \hat{b}^\dag_i \hat{b}_j \hat{b}^\dag_l \hat{b}_m \rangle &=& \delta_{j l}
\langle \hat{b}^\dag_i \hat{b}_m \rangle + \langle \hat{b}^\dag_i \hat{b}^\dag_l  \hat{b}_j \hat{b}_m \rangle
\nonumber \\
&=& 2\delta_{j l} \langle \hat{\sigma}^+_i \hat{\sigma}^-_m \rangle + (-1)^{\delta_{j l}}
\langle \hat{\sigma}^+_i \hat{\sigma}^-_j \hat{\sigma}^+_l \hat{\sigma}^-_m \rangle,\ \
\end{eqnarray}
where in the last step we have used Eq.\ (\ref{SpinComm}).

Next we note that last term in Eq.~\eqref{4pCorrHCB} can be rewritten as
\begin{eqnarray}
\label{4pCorrSpin} &&\langle \hat{\sigma}^+_i \hat{\sigma}^-_j \hat{\sigma}^+_l \hat{\sigma}^-_m \rangle
= \delta_{i j} \delta_{l m} + (-1)^{\delta_{i j}} \delta_{l m} \langle
\hat{\sigma}^-_j \hat{\sigma}^+_i \rangle \nonumber \\
&&\qquad + (-1)^{\delta_{l m}} \delta_{i j} \langle
\hat{\sigma}^-_m \hat{\sigma}^+_l \rangle
+ (-1)^{\delta_{i j}+\delta_{l m}} \delta_{i m}  \langle \hat{\sigma}^-_j
\hat{\sigma}^+_l \rangle \nonumber \\
&&\qquad +(-1)^{\delta_{i j}+\delta_{l m}+\delta_{i m}}
\langle \hat{\sigma}^-_j \hat{\sigma}^-_m \hat{\sigma}^+_i \hat{\sigma}^+_l \rangle \nonumber \\
&& \quad= \delta_{i j} \delta_{l m} + (-1)^{\delta_{i j}} \delta_{l m} G_{j i}
+ (-1)^{\delta_{l m}} \delta_{i j} G_{m l} \nonumber \\
&&\qquad + (-1)^{\delta_{i j}+\delta_{l m}} \delta_{i m} G_{j l}
+ (-1)^{\delta_{i j}+\delta_{l m}+\delta_{i m}} G_{j m i l},\qquad
\end{eqnarray}
where $G_{ijkl}=\langle \hat{\sigma}^-_i\hat{\sigma}^-_j\hat{\sigma}^+_k \hat{\sigma}^+_l\rangle$.
Note that all $G_{i j}$ can be obtained as described in the previous subsection.

Using the Jordan-Wigner transformation in Eq.\ (\ref{JWT}), the four-point Green's function
for the spin-1/2 system can be written as
\begin{eqnarray}
\label{Greens}
G_{ijkl} &=& \langle\Psi_{F}| \prod^{i-1}_{\alpha=1}e^{i\pi \hat{f}^{\dag}_{\alpha}\hat{f}^{}_{\alpha}} \hat{f}_i
\prod^{j-1}_{\beta=1}e^{i\pi \hat{f}^{\dag}_{\beta}\hat{f}^{}_{\beta}} \hat{f}_j \nonumber \\
&&\times \hat{f}^{\dag}_k \prod^{k-1}_{\gamma=1}e^{-i\pi \hat{f}^{\dag}_{\gamma}\hat{f}^{}_{\gamma}}
\hat{f}^{\dag}_l\prod^{l-1}_{\delta=1}e^{-i\pi \hat{f}^{\dag}_{\delta}\hat{f}^{}_{\delta}}| \Psi_{F}\rangle,
\end{eqnarray}
which using properties of Slater determinants, as described in
Refs.~\cite{rigol_muramatsu_04HCBa,rigol_muramatsu_05HCBb}, can be computed as
\begin{eqnarray}
\label{eq:4p}
G_{ijkl}=\det\left[ \left( {\bf P}^{ij} \right)^{\dag}{\bf P}^{kl}\right],
\end{eqnarray}
where $({\bf P}^{\alpha\beta})_{L,N+2}$, with $\alpha(\beta)=i,j,k,l$, is given by
\begin{eqnarray}
 P^{\alpha\beta}_{\varrho \kappa}= \left\{ \begin{array}{rl}
 -P^{\beta}_{\varrho \kappa} & \text{for } \varrho<    \alpha,\,\kappa=1,\ldots,N+1 \\
\,P^{\beta}_{\varrho \kappa} & \text{for } \varrho\geq \alpha,\,\kappa=1,\ldots,N+1 \\
  \delta_{\alpha\varrho} & \text{for } \kappa=N + 2
\end{array}\right.
\end{eqnarray}
and $({\bf P}^{\beta})_{L,N+1}$ is given by Eq.~\eqref{eq:matcomp}. This means that to
determine each element of the four-point Green's function we need to multiply a matrix of dimension
$(N+2)\times L$ by a matrix of dimension $L\times(N+2)$ [an operation that scales as
$(N+2)^2 L$] and then compute the determinant of the resulting $(N+2)\times (N+2)$
matrix [an operation that scales as $(N+2)^3$]. Finally, to compute the full four-point Green's
function, we need to calculate of the order of $L^4$ nonzero elements; that is, the total computation time
scales as $L^4[A(N+2)^2L+B(N+2)^3]$, with $A$ and $B$ being prefactors for matrix multiplications
and matrix determinants, respectively.

\section{Homogeneous case}\label{sec:periodic}

In this section we study the scaling of noise correlations in homogeneous chains. We should stress
that, for all hard-core-boson systems considered in the following, periodic boundary conditions are
always implemented; that is, for the equivalent fermionic Hamiltonians, periodic or antiperiodic conditions
are selected depending on the number of particles in the lattice.

\begin{figure}[!htb]
\begin{center}
\includegraphics[width=.48\textwidth]{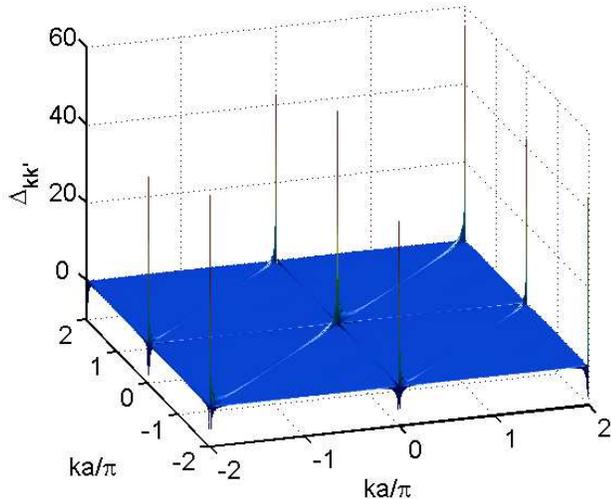}
\end{center}\vspace{-0.4cm}
\caption{Noise correlations as a function of $k$ and $k'$ for a
homogeneous system with 100 hard-core bosons in 200 lattice sites.}
\label{homofull}
\end{figure}

In Fig.\ \ref{homofull}, we show a typical noise correlation pattern for a strongly interacting
superfluid system. It was calculated in a periodic lattice with 200 sites at half-filling.
There are three features in that pattern that are apparent. First, a very large peak appears at
$k=k'=0$, reflecting the presence of quasicondensation in the system. Replicas of this peak
also appear at integer multiples of the reciprocal lattice vector $K$. Second, a line of maxima
can be found for $k=k'$ due to the usual bunching in bosonic systems. Finally, dips are seen along
the lines $k,0$ and $0,k'$, which are related to the quantum depletion in the
system. These features have been discussed in detail by Mathey {\it et al.}~\cite{mathey09} in the
more general context of Luttinger liquids, for which hard-core bosons correspond to a limiting case.

\begin{figure}[!htb]
\begin{center}
\includegraphics[width=.48\textwidth]{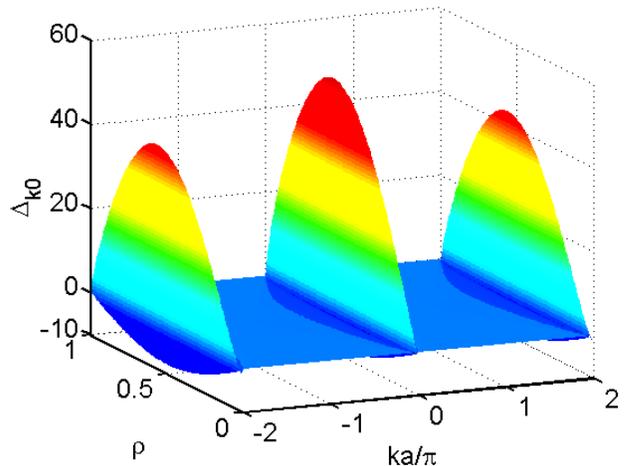}
\end{center}\vspace{-0.4cm}
\caption{Noise correlations for $k'=0$ as $k$ and $\rho$ are changed
for a system with $L=200$.} \label{homodens}
\end{figure}

As a function of the density $\rho=N/L$, the evolution of the noise correlations along the line $k,0$
is depicted in Fig.\ \ref{homodens}. The dips around the $k=0,\pm K$ peaks are more clearly seen in
Fig.\ \ref{homodens}. As noted in Ref.\ \cite{rey_satija_06a}, we find that the maximum value of $\Delta_{00}$
occurs for $\rho>0.5$, making evident the breakdown of the particle-hole symmetry for this observable.

\begin{figure}[!htb]
\begin{center}
\includegraphics[width=.41\textwidth]{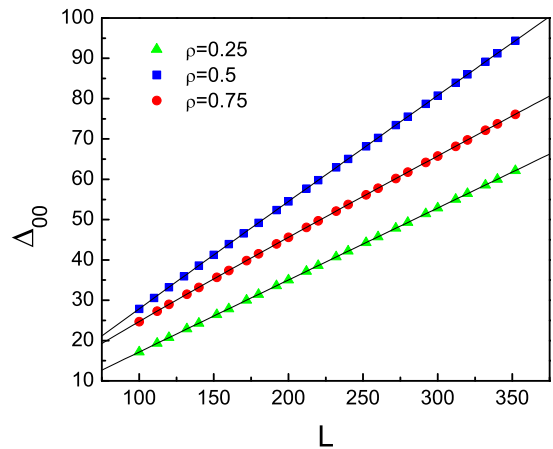}
\end{center}\vspace{-0.4cm}
\caption{Scaling of the noise correlations $\Delta_{0 0}$ for three
different densities, $\rho=$0.25, 0.5 and 0.75. The solid lines are
numerical fits in the form of Eq.\ (\ref{fit}) and the results (see
text) show a leading-order linear behavior.} \label{homosize}
\end{figure}

In what follows, we will focus on the scaling of the $\Delta_{00}$ for different densities.
For the $k=0$ peak of the momentum distribution function, it is well known that the power-law
decay of the one-particle correlations results in a $n_{k=0} \sim \sqrt{L}$ scaling \cite{lenard_64}.
In Fig.~\ref{homosize}, we show the scaling of $\Delta_{00}$ for three different densities in
our periodic systems.

To numerically find the scaling with system size, we assume that
\begin{equation}
\label{fit} \Delta_{0 0}=a L^x +b L^y,
\end{equation}
where $x$ and $y\ (y<x)$ describe the leading and subleading terms, respectively, and $a$ and $b$ are
coefficients that, together with $x$ and $y$, are determined by means of a numerical fit. The results
obtained for those four fitting parameters are given in Table~\ref{tab:homog}.
\begin{table}[!htb]
\caption{Fitting parameters for the homogeneous
case.}\label{tab:homog}
\begin{tabular}[t]{r@{\hspace{0.4cm}}r@{\hspace{0.4cm}}r@{\hspace{0.4cm}}r}\hline
&$\rho=0.25$    &$\rho=0.5$     &$\rho=0.75$ \\ \hline
$a$     &\qquad$0.17779(2)$     &$0.21(9)$  &$0.16(1)$ \\
$x$     &\qquad$1.00041(2)$     &$1.010(3)$     &$1.010(8)$  \\
$b$     &\qquad$-1.049(4)$  &$0.15(7)$  &$0.534(6)$  \\
$y$     &\qquad$-0.099(1)$  &$0.8(1)$   &$0.59(2)$  \\ \hline
\end{tabular}
\end{table}

Table \ref{tab:homog} shows that the leading term is essentially linear in all cases, while the exponent of the
power law of the subleading term does depend on the density and was found to be quite close to one
around half-filling. This means that in ultra-cold gas experiments one would need to reach large systems
sizes to be able to clearly observe the linear scaling of the noise correlation peaks around half-filling,
while this scaling would be more easily observed far away from half-filling.

\section{Period-two superlattice}\label{sec:period-2}

We now consider the case in which an additional lattice, with twice the periodicity of the original lattice,
is added to the system (a superlattice). In this case, the on-site potential in Hamiltonian (\ref{HamHCB})
has the form
\begin{equation}
\label{spPotential} V_i = V \cos(\pi i),
\end{equation}
with $V$ representing its strength. As discussed in Ref.~\cite{rousseau_arovas_06,rigol_muramatsu_06}, the effect
of a period-two superlattice is to open a gap of magnitude $V$ in the energy spectrum, splitting the
original band into two bands. As a result, besides the usual insulating phases at $\rho=0,1$, the half-filled system
in the ground state also exhibits insulating behavior. As opposed to the $\rho=0,1$ insulators, the $\rho=0.5$ (Mott)
insulator does exhibit nonzero density fluctuations and a finite correlation length.

\begin{figure}[!htb]
\begin{center}
\includegraphics[width=.48\textwidth]{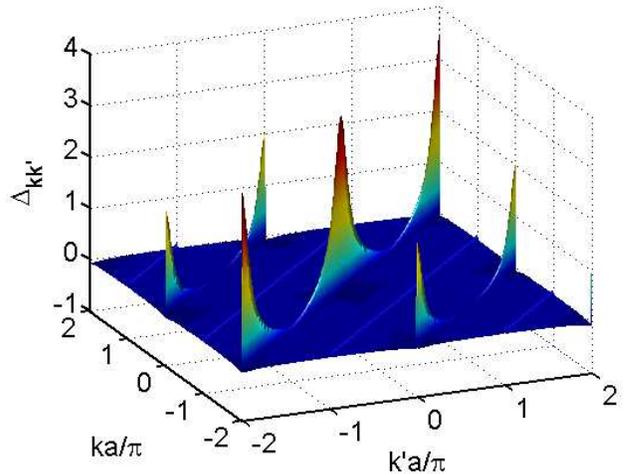}
\end{center}\vspace{-0.4cm}
\caption{Noise correlations for the fractional Mott phase in the
half-filled system in the presence of period-two superlattice for
$L=200$.} \label{spfull}
\end{figure}

Figure \ref{spfull} shows the noise correlation pattern for the
$\rho=0.5$ insulator with $V=1t$. Broad peaks can be clearly seen
along the line $k=k'$, and those are characteristic of the noise
correlations in the fractional Mott phase. They contrast with the
sharp peaks seen in the noise correlations of the superfluid regime
studied in the previous section. The suppressed height of the peaks
in Fig.~\ref{spfull} is a signature of the destruction of the
quasi-long-range coherence in the half-filled Mott system. At this
critical filling, the power-law decay of the one-particle
correlations present in the absence of the superlattice is
substituted by an exponential decay $\rho_{ij} \sim
\exp(-|i-j|/\xi)$, for which the correlation length $\xi$ was found
to be $\xi \sim 1/V$ for small values of $V$ ($V<t$) and $\xi \sim
1/\sqrt{V}$ for large values of $V$ ($V>t$)
\cite{rigol_muramatsu_06}. As long as the lattice sizes are
sufficiently large ($L\gg\xi$), the absence of quasi-long-range
coherence should be clearly observed in the scaling of the noise
correlations in those systems.

In the presence of the superlattice potential, additional features emerge in the noise correlations for
$k=k'\pm \pi/a$. Those can actually be used to distinguish the fractional insulator from the integer Mott insulator
as both have suppressed $\Delta_{00}$ peaks but only the former has a structure in the noise correlations for
$k=k'\pm \pi/a$.

\begin{figure}[!htb]
\begin{center}
\includegraphics[width=0.50\textwidth]{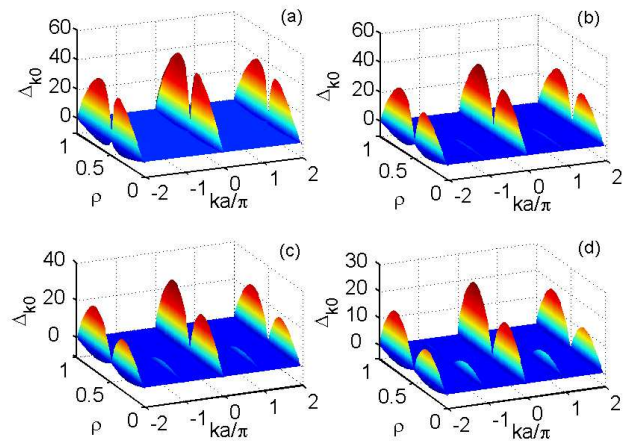}
\end{center}\vspace{-0.4cm}
\caption{Noise correlations $\Delta_{k 0}$ as a function of $k$ and
the density for superlattice systems with (a) $V=1t$, (b) $V=2t$,
(c) $V=3t$, and (d) $V=4t$. $L=200$ in all cases.} \label{spdens}
\end{figure}

In Fig.~\ref{spdens}, we present a unified view of the behavior of the noise correlations for different systems
with a superlattice potential. There we plot $\Delta_{k0}$ as a function of $\rho$ and $k$ for four different values
of $V$. Figure \ref{spdens} shows that as $V$ increases from $1t$ to $4t$, the intensity of the central peak decreases for
all fillings. However, the suppression is more dramatic around half-filling. The additional unique signature of the
presence of a superlattice potential is the structure that can be found at $ka=\pm\pi$. It is usually a positive
peak for densities below 0.5 and becomes a dip right after the density increases beyond the fractional filling
insulating phase. This peak-to-dip transition was discussed in detail by Rey
{\it et al.}~\cite{rey_satija_06b,rey_satija_06c}, where in the limit $V\rightarrow\infty$ one can show analytically
that $\Delta_{00}$ and $\Delta_{\pm\frac{\pi}{a}0}$ have different signs for $N=L/2+1$.

\begin{figure}[!htb]
\begin{center}
\includegraphics[width=0.41\textwidth]{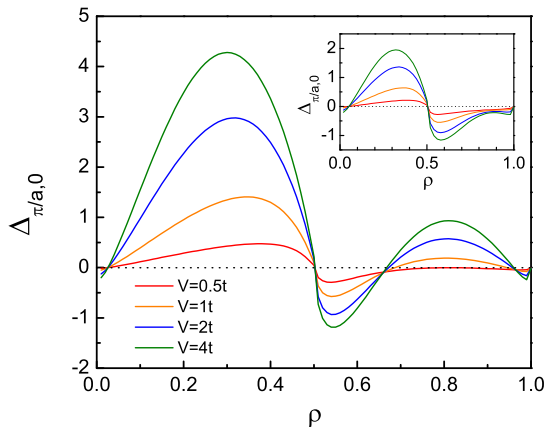}
\end{center}\vspace{-0.4cm}
\caption{The sublattice peak $\Delta_{\frac{\pi}{a}, 0}$ as a
function of $\rho$ for four different values of $V$ in systems with
200 sites. The inset shows the same quantity for systems with 100
sites.} \label{spExPeak}
\end{figure}

The behavior of $\Delta_{\frac{\pi}{a} 0}$ as a function of the density and for different values of $V$ can be
better seen in Fig.\ \ref{spExPeak}. Interestingly we find that, in addition to the peak to dip transition around
$\rho=0.5$, there are other dip to peak and peak to dip transitions for higher densities. Those are only apparent
for sufficiently large system sizes (beyond the ones studied in Refs.\ \cite{rey_satija_06b,rey_satija_06c}).
The inset in Fig.\ \ref{spExPeak} shows that for a smaller system size with only 100 sites
$\Delta_{\frac{\pi}{a}0}$ is always negative for $\rho>0.5$.

\begin{figure}[!htb]
\begin{center}
\includegraphics[width=0.41\textwidth]{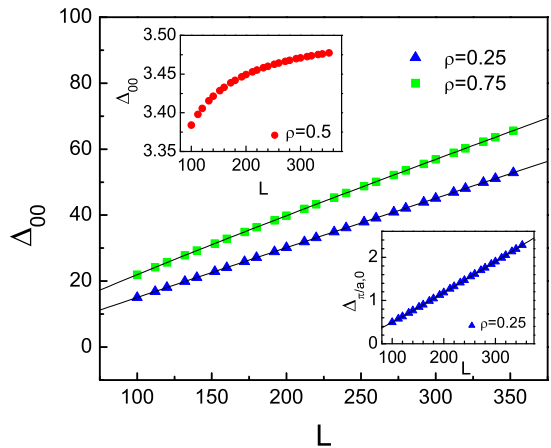}
\end{center}\vspace{-0.4cm}
\caption{Scaling of $\Delta_{0 0}$ for $\rho=0.25$ and
$\rho=0.75$ in systems with $V=1t$ and lattices with 200 sites.
The top inset shows results for the same systems in the fractional
($\rho=0.5$) Mott-insulating phase. The bottom inset shows scaling of the
sublattice peak $\Delta_{\frac{\pi}{a} 0}$ for $\rho=0.25$ in the
same systems. Solid lines are numerical fits to the results,
exhibiting a leading linear scaling with $L$ in all superfluid
cases.} \label{spsize}
\end{figure}

Now that the generic features of the noise correlations in a superlattice potential have been reviewed, we focus on
the scaling of the peaks with system size. In the fractional insulating regime, one expects that the exponential decay
of correlations should lead to a saturation of the noise correlation peaks. This is, indeed, what we find, and an example is depicted in the top inset in Fig.~\ref{spsize} for half-filled systems with $V=t$.

In the superfluid phases, on the other hand, it has been shown that one-particle correlations decay with exactly
the same power law as the homogeneous system \cite{rigol_muramatsu_06}. Hence, we expect to find the same leading
order scaling of $\Delta_{00}$ that was discussed for homogeneous systems in the previous section. This result can be seen in the main panel of Fig.~\ref{spsize}, and it can also be seen for the $\Delta_{\frac{\pi}{a}0}$ peak, for $\rho=0.25$, in the bottom inset.

Assuming the same scaling ansatz in Eq.~\eqref{fit}, but in the presence of the superlattice only used for the
superfluid phases, we obtain the values depicted in Table~\ref{tab:superlattice} for the fitting parameters
\begin{table}[!htb]
\caption{Fitting parameters for the superlattice
case.}\label{tab:superlattice}
\begin{tabular}[t]{rr@{\hspace{0.4cm}}rr}
\hline
&\multicolumn{2}{c}{\qquad$k=0$}    &$k=\pi/a$ \\
&$\rho=0.25$    &$\rho=0.75$    &$\rho=0.25$ \\ \hline
$a$ &\qquad$0.15928(6)$     &$0.090(1)$     &\qquad $0.00716(2)$ \\
$x$ &\qquad$0.99023(6)$     &$1.045(1)$     &\qquad $1.0133(3)$  \\
$b$ &\qquad$-5.2(3)$    &$0.5793(5)$    &\qquad $-0.03891(9)$  \\
$y$ &\qquad$-0.71(1)$   &$0.636(2)$     &\qquad $0.417(1)$  \\ \hline
\end{tabular}
\end{table}

As expected, we find that the leading terms of the noise correlation peaks are also of order $L$ for both
$\Delta_{0 0}$ and $\Delta_{\frac{\pi}{a}0}$. Similar to the homogeneous case, the leading linear scaling
of those peaks is better seen at low densities where finite-size effects have been found to be smaller because the subleading term has a much slower scaling with system size than the leading term.

\section{Disordered system}\label{sec:disorder}

The disordered case is simulated by a random on-site potential of the form
\begin{equation}
\label{disoPotential} V_i = \delta \epsilon_i,
\end{equation}
where $\delta$ represents the strength of disorder and
\{$\epsilon_i$\} are a set of random numbers between -1 and 1
selected with a uniform probability distribution. For our disorder
calculations we usually average over between 128 and 256 disorder
realizations.

\begin{figure}[!htb]
\begin{center}
\includegraphics[width=0.41\textwidth]{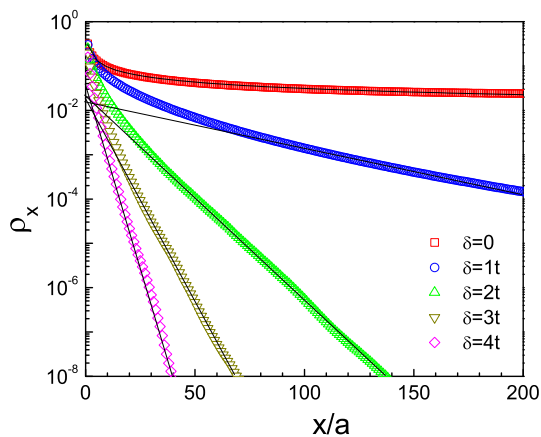}
\end{center}\vspace{-0.4cm}
\caption{The decay of one-particle density matrices in half-filled
systems with $L=500$, characterized by different disorder strength.
The disorder averaging is performed over 128 realizations for all
$\delta\neq0$ cases. Solid lines depict exponential decay, except
for the homogeneous ($\delta=0$) system, where a solid line depicts the known power
law $\sqrt{x}$. Note the log-linear scale.} \label{disooffdiag}
\end{figure}

For one-dimensional noninteracting fermionic systems, the presence of disorder is known to lead to Anderson
localization. This is a phase in which correlations decay exponentially while the system remains compressible; that is, no gap is present in the energy spectrum. Since hard-core bosons can be mapped to noninteracting
fermions, the same is known to be true for the former. We should note that despite the fact that the one-particle
correlations of hard-core bosons are in general different from those of noninteracting fermions, they also decay
exponentially. This is shown in Fig.\ \ref{disooffdiag}, where we present $\rho_x$ (with $x=|i-j|$) for systems
with different disorder strengths. One should note that the exponential decay always sets in beyond a certain
distance, which decreases as the strength of the disorder increases; that is, small systems with weak disorder
may behave as superfluids.

\begin{figure}[!htb]
\begin{center}
\includegraphics[width=0.5\textwidth]{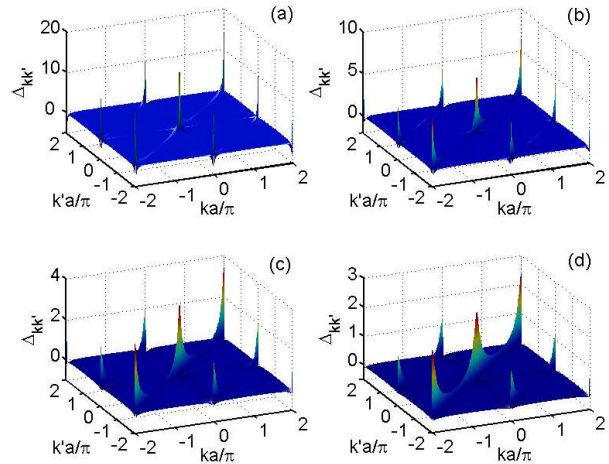}
\end{center}\vspace{-0.4cm}
\caption{Disorder-averaged noise correlations as a function of $k$
and $k'$ for systems with different disorder strength $\delta=$(a) $1t$,\
(b) $2t$,\ (c) $3t$, and (d) $4t$. $N=50$ and $L=100$ for all cases and the average was
performed over 128 disorder realizations.} \label{disofull}
\end{figure}

From the previous discussion and the results in Fig.~\ref{disooffdiag} one expects that, for any given system size,
the height of the noise correlation peaks should decrease with increasing disorder strength and the peaks should become broader. This can be seen in Fig.\ \ref{disofull}, where we show the noise correlations $\Delta_{k k'}$ for
four different disordered strengths in systems with 100 sites. The pattern for the $\delta=1$ case resembles
that of a homogeneous superfluid system (Fig.\ \ref{homofull}), while for larger values of $\delta$ they display
more similarities with the fractional Mott-insulating phase in the half-filled superlattice systems, with a clear
broadening of the peaks at $k=k'$. (Of course, no additional feature appear for $k=k'\pm\pi/a$ in the disordered case.)

\begin{figure}[!htb]
\begin{center}
\includegraphics[width=0.41\textwidth]{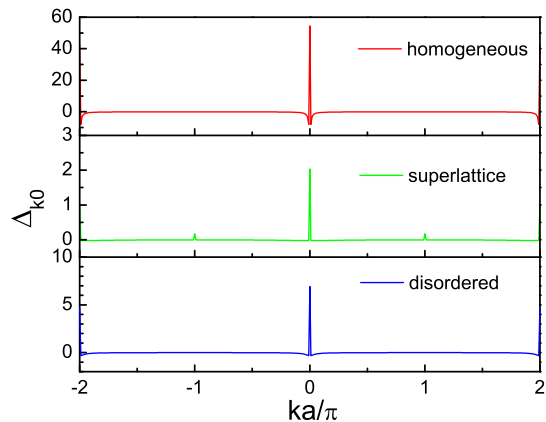}
\end{center}\vspace{-0.4cm}
\caption{Noise correlations with fixed $k'=0$ for three half-filled
systems with $L=200$. The superfluid phase, the fractional Mott
phase, and the Anderson-glass phase are associated with the
homogeneous, period-two superlattice ($V=2t$), and disordered
($\delta=2t$) cases, respectively. The average is performed over 256
disorder realizations.} \label{peakcontrast}
\end{figure}

A comparison between the cross-sectional view (for $k'=0$) of the noise correlations in all three phases discussed
previously, namely, the superfluid, fractional Mott, and glassy phases, is shown in Fig.\ \ref{peakcontrast}. This
comparison makes evident (i) the suppression of the $\Delta_{0 0}$ peak in the fractional Mott and glassy phases,
(ii) the fact that the two insulating phases can be distinguished by the superlattice-induced features at
$k=k'\pm \pi$, and (iii) that the disordered Anderson-glass and the superfluid phase exhibit the same satellite
dips accompanying the $\Delta_{0 0}$ peaks, while the dips vanish rather quickly in the fractional Mott phase.

\begin{figure}[!htb]
\begin{center}
\includegraphics[width=0.41\textwidth]{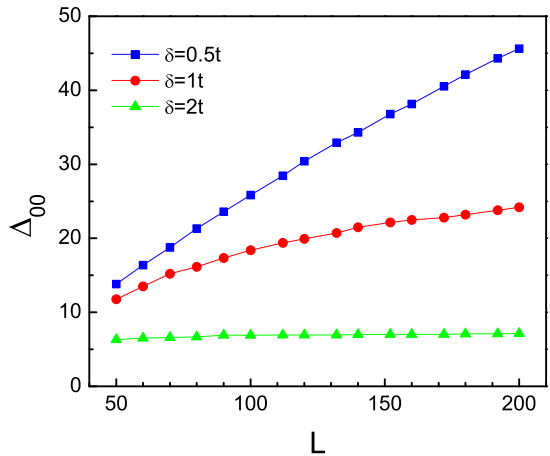}
\end{center}\vspace{-0.4cm}
\caption{Scaling of the noise correlations $\Delta_{0 0}$ for three
different values of $\delta$ in half-filled systems.}
\label{disosize}
\end{figure}

Similar to the behavior of the fractional Mott phase, one also expects that as the system size increases for
any nonzero value of the disorder strength, the $\Delta_{0 0}$ will saturate to a size-independent value that will
only be a function of the density and the disorder strength. This behavior is shown in Fig.~\ref{disosize} for
three different values of the disorder strength and for systems with up to 200 sites.

Finally, we study how the $\Delta_{0 0}$ peak in the noise correlations behaves as a function of the
disorder strength for a fixed size of the lattice. Since for $\delta=0$ we have already shown that such a peak
diverges with system size, in the following we analyze how $\Delta_{0 0}$ decreases as the disorder strength
increases.

\begin{figure}[!htb]
\begin{center}
\includegraphics[width=0.41\textwidth]{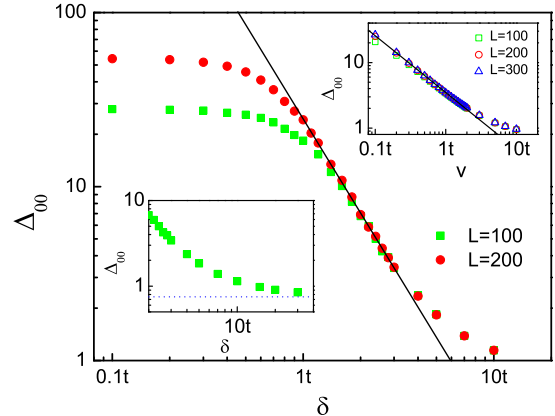}
\end{center}\vspace{-0.4cm}
\caption{Noise correlation peaks $\Delta_{0 0}$ as a
function of disorder strength in two half-filled disordered systems
with $L=$100 and 200, respectively. The solid line shows a power-law
fit $\Delta_{0 0}\sim \delta^{-\gamma}$, with $\gamma=1.78(2)$ in
the range from $\delta=t$ to $3t$ for $L=200$. The top inset shows
$\Delta_{0 0}$ as a function of $V$ in three half-filled period-two
superlattices. The solid line depicts a power law with an exponent
$\gamma=0.874(5)$. The bottom inset shows the asymptotic behavior for large
values of $\delta$; the dotted line marks the analytical result in
the limit of infinite disorder.} \label{disoamp}
\end{figure}

In Fig.\ \ref{disoamp}, we show $\Delta_{0 0}$ as a function of
$\delta$ for two different system sizes. Three different regimes can
be clearly identified. (i) For small values of
$\delta$, $\Delta_{0 0}$ approximately stays constant with the
increase of $\delta$, which can be understood to be a consequence of
a correlation length that exceeds the system size. As seen in Fig.\ref{disoamp}, that region decreases as the system size increases. (ii) As $\delta$ increases even further, a power-law decay develops in
$\Delta_{0 0}$, and the region over which such a power law can be
seen increases with system size as regime (i) is suppressed. In
our fits, we find the power law $\Delta_{0 0}\sim \delta^{-\gamma}$
to have an exponent $\gamma\sim 1.78(2)$, but it is still influenced by some
finite-size effects. In order to gain further understanding of the power-law
decay of the height of this noise peak in insulating phases, we have studied the behavior
of $\Delta_{0 0}$ vs $V$ in the fractional Mott phase in a superlattice, for which we
can study larger systems sizes. We find that $\Delta_{0 0}\sim V^{-\gamma}$
with an exponent of $0.874(5)$, which is different from the one for
the disordered system.
These results clearly show that the power-law decay of $\Delta_{0 0}$
as one enters an insulating phase depends on the perturbation creating
the insulator, i.e., it is not universal. (iii) Finally, for very strong disorder,
$\Delta_{0 0}$ saturates to a nonzero value. This asymptotic
behavior is found to agree with the analytical value in the $\delta
\rightarrow \infty$ limit, computed using
\begin{equation}
\label{infdiso} \Delta_{0 0} = \rho(\rho+1),
\end{equation}
which was derived by Rey \textit{et al.} \cite{rey_satija_06b}. Equation (\ref{infdiso}) shows
that $\Delta_{0 0}$ only depends on the
density and also makes explicit the absence of particle-hole symmetry
for this observable in hard-core-boson systems. This third regime is
robust against the disorder variance, something that follows from
the fact that the correlation length is of the order of or smaller
than the lattice spacing $a$.

\section{Conclusions}\label{sec:Conclusions}

We have implemented an exact approach to numerically compute the noise correlations for hard-core boson
in one-dimensional lattices. For that purpose, we have extended to four-point correlations the recipe for calculating
two-point correlations introduced in Refs.~\cite{rigol_muramatsu_04HCBa,rigol_muramatsu_05HCBb}. Our approach has
a polynomial time scaling that is more efficient than the straightforward application of Wicks theorem, and can be
easily extended to study noise correlations in nonequilibrium systems.

We have applied this approach to  study the scaling of noise
correlations in three different phases that appear in homogeneous
systems and in the presence of two different background potentials.
We have shown that in the superfluid phase, the noise correlation
peaks $\Delta_{0 0}$ exhibit a leading linear behavior $\sim L$,
independent of the density and of the presence of a superlattice
potential. The subleading term was found to be strongly dependent on
the latter two. On the other hand, the fractional Mott and
Anderson-glass insulating phases exhibit an asymptotic value, which
is independent of system size and only depends on the density and
the strength of the potentials creating such phases. This behavior
was expected and manifests the absence of quasi-long-range order in
these two phases.

In the period-two superlattice, we have also found various peak-to-dip and dip-to-peak transitions that were not
observed in previous studies with smaller system sizes, something that demonstrates the importance of finite-size
effects in the noise correlations and the need for approaches that allow one to study very large system sizes.

Finally, we have shown that in the disordered system
(fractional Mott phase), the decrease of the $\Delta_{0 0}$ peak
with increasing disorder strength (superlattice strength) exhibits
a region with a power-law decay $\Delta_{0 0}\sim
\delta(V)^{-\gamma}$, with a nonuniversal value of exponent $\gamma$
that depends on the kind of perturbation creating the insulator.

\begin{acknowledgments}
This work was supported by the U.S. Office of Naval Research under Grant No.~N000140910966 and by the
National Science Foundation under Grant No.~PHY05-51164. We are grateful to Indubala I. Satija
for useful discussions.
\end{acknowledgments}


\begin{thebibliography}{39}
\expandafter\ifx\csname natexlab\endcsname\relax\def\natexlab#1{#1}\fi
\expandafter\ifx\csname bibnamefont\endcsname\relax
  \def\bibnamefont#1{#1}\fi
\expandafter\ifx\csname bibfnamefont\endcsname\relax
  \def\bibfnamefont#1{#1}\fi
\expandafter\ifx\csname citenamefont\endcsname\relax
  \def\citenamefont#1{#1}\fi
\expandafter\ifx\csname url\endcsname\relax
  \def\url#1{\texttt{#1}}\fi
\expandafter\ifx\csname urlprefix\endcsname\relax\def\urlprefix{URL }\fi
\providecommand{\bibinfo}[2]{#2}
\providecommand{\eprint}[2][]{\url{#2}}

\bibitem[{\citenamefont{Bloch et~al.}(2008)\citenamefont{Bloch, Dalibard, and
  Zwerger}}]{bloch_dalibard_review_08}
\bibinfo{author}{\bibfnamefont{I.}~\bibnamefont{Bloch}},
  \bibinfo{author}{\bibfnamefont{J.}~\bibnamefont{Dalibard}}, \bibnamefont{and}
  \bibinfo{author}{\bibfnamefont{W.}~\bibnamefont{Zwerger}},
  \bibinfo{journal}{Rev. Mod. Phys.} \textbf{\bibinfo{volume}{80}},
  \bibinfo{pages}{885} (\bibinfo{year}{2008}).

\bibitem[{\citenamefont{Schreck et~al.}(2001)\citenamefont{Schreck, Khaykovich,
  Corwin, Ferrari, Bourdel, Cubizolles, and Salomon}}]{schreck_khaykovich_01}
\bibinfo{author}{\bibfnamefont{F.}~\bibnamefont{Schreck}},
  \bibinfo{author}{\bibfnamefont{L.}~\bibnamefont{Khaykovich}},
  \bibinfo{author}{\bibfnamefont{K.~L.} \bibnamefont{Corwin}},
  \bibinfo{author}{\bibfnamefont{G.}~\bibnamefont{Ferrari}},
  \bibinfo{author}{\bibfnamefont{T.}~\bibnamefont{Bourdel}},
  \bibinfo{author}{\bibfnamefont{J.}~\bibnamefont{Cubizolles}},
  \bibnamefont{and} \bibinfo{author}{\bibfnamefont{C.}~\bibnamefont{Salomon}},
  \bibinfo{journal}{Phys. Rev. Lett.} \textbf{\bibinfo{volume}{87}},
  \bibinfo{pages}{080403} (\bibinfo{year}{2001}).

\bibitem[{\citenamefont{G\"orlitz et~al.}(2001)\citenamefont{G\"orlitz, Vogels,
  Leanhardt, Raman, Gustavson, Abo-Shaeer, Chikkatur, Gupta, Inouye, Rosenband
  et~al.}}]{gorlitz_vogels_01}
\bibinfo{author}{\bibfnamefont{A.}~\bibnamefont{G\"orlitz}},
  \bibinfo{author}{\bibfnamefont{J.~M.} \bibnamefont{Vogels}},
  \bibinfo{author}{\bibfnamefont{A.~E.} \bibnamefont{Leanhardt}},
  \bibinfo{author}{\bibfnamefont{C.}~\bibnamefont{Raman}},
  \bibinfo{author}{\bibfnamefont{T.~L.} \bibnamefont{Gustavson}},
  \bibinfo{author}{\bibfnamefont{J.~R.} \bibnamefont{Abo-Shaeer}},
  \bibinfo{author}{\bibfnamefont{A.~P.} \bibnamefont{Chikkatur}},
  \bibinfo{author}{\bibfnamefont{S.}~\bibnamefont{Gupta}},
  \bibinfo{author}{\bibfnamefont{S.}~\bibnamefont{Inouye}},
  \bibinfo{author}{\bibfnamefont{T.}~\bibnamefont{Rosenband}},
  \bibnamefont{and} \bibinfo{author}{\bibfnamefont{W.}~\bibnamefont{Ketterle}},
  \bibinfo{journal}{Phys. Rev. Lett.}
  \textbf{\bibinfo{volume}{87}}, \bibinfo{pages}{130402}
  (\bibinfo{year}{2001}).

\bibitem[{\citenamefont{Greiner et~al.}(2001)\citenamefont{Greiner, Bloch,
  Mandel, H\"ansch, and Esslinger}}]{greiner_bloch_01}
\bibinfo{author}{\bibfnamefont{M.}~\bibnamefont{Greiner}},
  \bibinfo{author}{\bibfnamefont{I.}~\bibnamefont{Bloch}},
  \bibinfo{author}{\bibfnamefont{O.}~\bibnamefont{Mandel}},
  \bibinfo{author}{\bibfnamefont{T.~W.} \bibnamefont{H\"ansch}},
  \bibnamefont{and}
  \bibinfo{author}{\bibfnamefont{T.}~\bibnamefont{Esslinger}},
  \bibinfo{journal}{Phys. Rev. Lett.} \textbf{\bibinfo{volume}{87}},
  \bibinfo{pages}{160405} (\bibinfo{year}{2001}).

\bibitem[{\citenamefont{Moritz et~al.}(2003)\citenamefont{Moritz, St\"oferle,
  K\"ohl, and Esslinger}}]{moritz_stoferle_03}
\bibinfo{author}{\bibfnamefont{H.}~\bibnamefont{Moritz}},
  \bibinfo{author}{\bibfnamefont{T.}~\bibnamefont{St\"oferle}},
  \bibinfo{author}{\bibfnamefont{M.}~\bibnamefont{K\"ohl}}, \bibnamefont{and}
  \bibinfo{author}{\bibfnamefont{T.}~\bibnamefont{Esslinger}},
  \bibinfo{journal}{Phys. Rev. Lett.} \textbf{\bibinfo{volume}{91}},
  \bibinfo{pages}{250402} (\bibinfo{year}{2003}).

\bibitem[{\citenamefont{St\"oferle et~al.}(2004)\citenamefont{St\"oferle,
  Moritz, Schori, K\"ohl, and Esslinger}}]{stoferle_moritz_04}
\bibinfo{author}{\bibfnamefont{T.}~\bibnamefont{St\"oferle}},
  \bibinfo{author}{\bibfnamefont{H.}~\bibnamefont{Moritz}},
  \bibinfo{author}{\bibfnamefont{C.}~\bibnamefont{Schori}},
  \bibinfo{author}{\bibfnamefont{M.}~\bibnamefont{K\"ohl}}, \bibnamefont{and}
  \bibinfo{author}{\bibfnamefont{T.}~\bibnamefont{Esslinger}},
  \bibinfo{journal}{Phys. Rev. Lett.} \textbf{\bibinfo{volume}{92}},
  \bibinfo{pages}{130403} (\bibinfo{year}{2004}).

\bibitem[{\citenamefont{Paredes et~al.}(2004)\citenamefont{Paredes, Widera,
  Murg, Mandel, F\"olling, Cirac, Shlyapnikov, H\"ansch, and
  Bloch}}]{paredes_widera_04}
\bibinfo{author}{\bibfnamefont{B.}~\bibnamefont{Paredes}},
  \bibinfo{author}{\bibfnamefont{A.}~\bibnamefont{Widera}},
  \bibinfo{author}{\bibfnamefont{V.}~\bibnamefont{Murg}},
  \bibinfo{author}{\bibfnamefont{O.}~\bibnamefont{Mandel}},
  \bibinfo{author}{\bibfnamefont{S.}~\bibnamefont{F\"olling}},
  \bibinfo{author}{\bibfnamefont{I.}~\bibnamefont{Cirac}},
  \bibinfo{author}{\bibfnamefont{G.~V.} \bibnamefont{Shlyapnikov}},
  \bibinfo{author}{\bibfnamefont{T.~W.} \bibnamefont{H\"ansch}},
  \bibnamefont{and} \bibinfo{author}{\bibfnamefont{I.}~\bibnamefont{Bloch}},
  \bibinfo{journal}{Nature (London)} \textbf{\bibinfo{volume}{429}},
  \bibinfo{pages}{277} (\bibinfo{year}{2004}).

\bibitem[{\citenamefont{Kinoshita et~al.}(2004)\citenamefont{Kinoshita, Wenger,
  and Weiss}}]{kinoshita_wenger_04}
\bibinfo{author}{\bibfnamefont{T.}~\bibnamefont{Kinoshita}},
  \bibinfo{author}{\bibfnamefont{T.}~\bibnamefont{Wenger}}, \bibnamefont{and}
  \bibinfo{author}{\bibfnamefont{D.~S.} \bibnamefont{Weiss}},
  \bibinfo{journal}{Science} \textbf{\bibinfo{volume}{305}},
  \bibinfo{pages}{1125} (\bibinfo{year}{2004}).

\bibitem[{\citenamefont{Kinoshita et~al.}(2005)\citenamefont{Kinoshita, Wenger,
  and Weiss}}]{kinoshita_wenger_05}
\bibinfo{author}{\bibfnamefont{T.}~\bibnamefont{Kinoshita}},
  \bibinfo{author}{\bibfnamefont{T.}~\bibnamefont{Wenger}}, \bibnamefont{and}
  \bibinfo{author}{\bibfnamefont{D.~S.} \bibnamefont{Weiss}},
  \bibinfo{journal}{Phys. Rev. Lett.} \textbf{\bibinfo{volume}{95}},
  \bibinfo{pages}{190406} (\bibinfo{year}{2005}).

\bibitem[{\citenamefont{Trebbia et~al.}(2006)\citenamefont{Trebbia, Esteve,
  Westbrook, and Bouchoule}}]{trebbia_esteve_06}
\bibinfo{author}{\bibfnamefont{J.-B.} \bibnamefont{Trebbia}},
  \bibinfo{author}{\bibfnamefont{J.}~\bibnamefont{Esteve}},
  \bibinfo{author}{\bibfnamefont{C.~I.} \bibnamefont{Westbrook}},
  \bibnamefont{and}
  \bibinfo{author}{\bibfnamefont{I.}~\bibnamefont{Bouchoule}},
  \bibinfo{journal}{Phys. Rev. Lett.} \textbf{\bibinfo{volume}{97}},
  \bibinfo{pages}{250403} (\bibinfo{year}{2006}).

\bibitem[{\citenamefont{Hofferberth et~al.}(2007)\citenamefont{Hofferberth,
  Lesanovsky, Fischer, Schumm, and Schmiedmayer}}]{hofferberth_lesanovsky_07}
\bibinfo{author}{\bibfnamefont{S.}~\bibnamefont{Hofferberth}},
  \bibinfo{author}{\bibfnamefont{I.}~\bibnamefont{Lesanovsky}},
  \bibinfo{author}{\bibfnamefont{B.}~\bibnamefont{Fischer}},
  \bibinfo{author}{\bibfnamefont{T.}~\bibnamefont{Schumm}}, \bibnamefont{and}
  \bibinfo{author}{\bibfnamefont{J.}~\bibnamefont{Schmiedmayer}},
  \bibinfo{journal}{Nature (London)} \textbf{\bibinfo{volume}{449}},
  \bibinfo{pages}{324} (\bibinfo{year}{2007}).

\bibitem[{\citenamefont{van Amerongen et~al.}(2008)\citenamefont{van Amerongen,
  van Es, Wicke, Kheruntsyan, and van Druten}}]{amerongen_es_08}
\bibinfo{author}{\bibfnamefont{A.~H.} \bibnamefont{van Amerongen}},
  \bibinfo{author}{\bibfnamefont{J.~J.~P.} \bibnamefont{van Es}},
  \bibinfo{author}{\bibfnamefont{P.}~\bibnamefont{Wicke}},
  \bibinfo{author}{\bibfnamefont{K.~V.} \bibnamefont{Kheruntsyan}},
  \bibnamefont{and} \bibinfo{author}{\bibfnamefont{N.~J.} \bibnamefont{van
  Druten}}, \bibinfo{journal}{Phys. Rev. Lett.} \textbf{\bibinfo{volume}{100}},
  \bibinfo{pages}{090402} (\bibinfo{year}{2008}).

\bibitem[{\citenamefont{Girardeau}(1960)}]{girardeau_60}
\bibinfo{author}{\bibfnamefont{M.}~\bibnamefont{Girardeau}},
  \bibinfo{journal}{J. Math. Phys.} \textbf{\bibinfo{volume}{1}},
  \bibinfo{pages}{516} (\bibinfo{year}{1960}).

\bibitem[{\citenamefont{Lieb et~al.}(1961)\citenamefont{Lieb, Shultz, and
  Mattis}}]{lieb_61}
\bibinfo{author}{\bibfnamefont{E.}~\bibnamefont{Lieb}},
  \bibinfo{author}{\bibfnamefont{T.}~\bibnamefont{Shultz}}, \bibnamefont{and}
  \bibinfo{author}{\bibfnamefont{D.}~\bibnamefont{Mattis}},
  \bibinfo{journal}{Ann. Phys. (N.Y.)} \textbf{\bibinfo{volume}{16}},
  \bibinfo{pages}{406} (\bibinfo{year}{1961}).

\bibitem[{\citenamefont{Lenard}(1964)}]{lenard_64}
\bibinfo{author}{\bibfnamefont{A.}~\bibnamefont{Lenard}}, \bibinfo{journal}{J.
  Math. Phys.} \textbf{\bibinfo{volume}{5}}, \bibinfo{pages}{930}
  (\bibinfo{year}{1964}).

\bibitem[{\citenamefont{McCoy}(1968)}]{mccoy_68}
\bibinfo{author}{\bibfnamefont{B.~M.} \bibnamefont{McCoy}},
  \bibinfo{journal}{Phys. Rev.} \textbf{\bibinfo{volume}{173}},
  \bibinfo{pages}{531} (\bibinfo{year}{1968}).

\bibitem[{\citenamefont{Vaidya and Tracy}(1978)}]{vaidya_tracy_78}
\bibinfo{author}{\bibfnamefont{H.~G.} \bibnamefont{Vaidya}} \bibnamefont{and}
  \bibinfo{author}{\bibfnamefont{C.~A.} \bibnamefont{Tracy}},
  \bibinfo{journal}{Phys. Lett. A} \textbf{\bibinfo{volume}{68}},
  \bibinfo{pages}{378 } (\bibinfo{year}{1978}).

\bibitem[{\citenamefont{Vaidya and Tracy}(1979)}]{vaidya_tracy_79a}
\bibinfo{author}{\bibfnamefont{H.~G.} \bibnamefont{Vaidya}} \bibnamefont{and}
  \bibinfo{author}{\bibfnamefont{C.~A.} \bibnamefont{Tracy}},
  \bibinfo{journal}{Phys. Rev. Lett.} \textbf{\bibinfo{volume}{42}},
  \bibinfo{pages}{3} (\bibinfo{year}{1979}).

\bibitem[{\citenamefont{Jimbo et~al.}(1980)\citenamefont{Jimbo, Miwa, M{\^o}ri,
  and Sato}}]{jimbo_80}
\bibinfo{author}{\bibfnamefont{M.}~\bibnamefont{Jimbo}},
  \bibinfo{author}{\bibfnamefont{T.}~\bibnamefont{Miwa}},
  \bibinfo{author}{\bibfnamefont{Y.}~\bibnamefont{M{\^o}ri}}, \bibnamefont{and}
  \bibinfo{author}{\bibfnamefont{M.}~\bibnamefont{Sato}},
  \bibinfo{journal}{Phys. D} \textbf{\bibinfo{volume}{1}},
  \bibinfo{pages}{80} (\bibinfo{year}{1980}).

\bibitem[{\citenamefont{Gangardt}(2004)}]{gangardt_04}
\bibinfo{author}{\bibfnamefont{D.~M.} \bibnamefont{Gangardt}},
  \bibinfo{journal}{J. Phys. A} \textbf{\bibinfo{volume}{37}},
  \bibinfo{pages}{9335} (\bibinfo{year}{2004}).

\bibitem[{\citenamefont{Girardeau et~al.}(2001)\citenamefont{Girardeau, Wright,
  and Triscari}}]{girardeau_wright_01}
\bibinfo{author}{\bibfnamefont{M.~D.} \bibnamefont{Girardeau}},
  \bibinfo{author}{\bibfnamefont{E.~M.} \bibnamefont{Wright}},
  \bibnamefont{and} \bibinfo{author}{\bibfnamefont{J.~M.}
  \bibnamefont{Triscari}}, \bibinfo{journal}{Phys. Rev. A}
  \textbf{\bibinfo{volume}{63}}, \bibinfo{pages}{033601}
  (\bibinfo{year}{2001}).

\bibitem[{\citenamefont{Lapeyre et~al.}(2002)\citenamefont{Lapeyre, Girardeau,
  and Wright}}]{lapeyre_girardeau_02}
\bibinfo{author}{\bibfnamefont{G.~J.} \bibnamefont{Lapeyre}},
  \bibinfo{author}{\bibfnamefont{M.~D.} \bibnamefont{Girardeau}},
  \bibnamefont{and} \bibinfo{author}{\bibfnamefont{E.~M.}
  \bibnamefont{Wright}}, \bibinfo{journal}{Phys. Rev. A}
  \textbf{\bibinfo{volume}{66}}, \bibinfo{pages}{023606}
  (\bibinfo{year}{2002}).

\bibitem[{\citenamefont{Papenbrock}(2003)}]{papenbrock_03}
\bibinfo{author}{\bibfnamefont{T.}~\bibnamefont{Papenbrock}},
  \bibinfo{journal}{Phys. Rev. A} \textbf{\bibinfo{volume}{67}},
  \bibinfo{pages}{041601(R)} (\bibinfo{year}{2003}).

\bibitem[{\citenamefont{Forrester et~al.}(2003)\citenamefont{Forrester,
  Frankel, Garoni, and Witte}}]{forrester_03b}
\bibinfo{author}{\bibfnamefont{P.~J.} \bibnamefont{Forrester}},
  \bibinfo{author}{\bibfnamefont{N.~E.} \bibnamefont{Frankel}},
  \bibinfo{author}{\bibfnamefont{T.~M.} \bibnamefont{Garoni}},
  \bibnamefont{and} \bibinfo{author}{\bibfnamefont{N.~S.} \bibnamefont{Witte}},
  \bibinfo{journal}{Phys. Rev. A} \textbf{\bibinfo{volume}{67}},
  \bibinfo{pages}{043607} (\bibinfo{year}{2003}).

\bibitem[{\citenamefont{Rigol and Muramatsu}(2004)}]{rigol_muramatsu_04HCBa}
\bibinfo{author}{\bibfnamefont{M.}~\bibnamefont{Rigol}} \bibnamefont{and}
  \bibinfo{author}{\bibfnamefont{A.}~\bibnamefont{Muramatsu}},
  \bibinfo{journal}{Phys. Rev. A} \textbf{\bibinfo{volume}{70}},
  \bibinfo{pages}{031603(R)} (\bibinfo{year}{2004}).

\bibitem[{\citenamefont{Rigol and Muramatsu}(2005)}]{rigol_muramatsu_05HCBb}
\bibinfo{author}{\bibfnamefont{M.}~\bibnamefont{Rigol}} \bibnamefont{and}
  \bibinfo{author}{\bibfnamefont{A.}~\bibnamefont{Muramatsu}},
  \bibinfo{journal}{Phys. Rev. A} \textbf{\bibinfo{volume}{72}},
  \bibinfo{pages}{013604} (\bibinfo{year}{2005}).

\bibitem[{\citenamefont{Rigol}(2005)}]{rigol_05}
\bibinfo{author}{\bibfnamefont{M.}~\bibnamefont{Rigol}},
  \bibinfo{journal}{Phys. Rev. A} \textbf{\bibinfo{volume}{72}},
  \bibinfo{pages}{063607} (\bibinfo{year}{2005}).

\bibitem[{\citenamefont{Altman et~al.}(2004)\citenamefont{Altman, Demler, and
  Lukin}}]{altman_demler_04}
\bibinfo{author}{\bibfnamefont{E.}~\bibnamefont{Altman}},
  \bibinfo{author}{\bibfnamefont{E.}~\bibnamefont{Demler}}, \bibnamefont{and}
  \bibinfo{author}{\bibfnamefont{M.~D.} \bibnamefont{Lukin}},
  \bibinfo{journal}{Phys. Rev. A} \textbf{\bibinfo{volume}{70}},
  \bibinfo{pages}{013603} (\bibinfo{year}{2004}).

\bibitem[{\citenamefont{F\"olling et~al.}(2005)\citenamefont{F\"olling,
  Gerbier, Widera, Mandel, Gericke, and Bloch}}]{folling_gerbier_05}
\bibinfo{author}{\bibfnamefont{S.}~\bibnamefont{F\"olling}},
  \bibinfo{author}{\bibfnamefont{F.}~\bibnamefont{Gerbier}},
  \bibinfo{author}{\bibfnamefont{A.}~\bibnamefont{Widera}},
  \bibinfo{author}{\bibfnamefont{O.}~\bibnamefont{Mandel}},
  \bibinfo{author}{\bibfnamefont{T.}~\bibnamefont{Gericke}}, \bibnamefont{and}
  \bibinfo{author}{\bibfnamefont{I.}~\bibnamefont{Bloch}},
  \bibinfo{journal}{Nature (London)} \textbf{\bibinfo{volume}{434}},
  \bibinfo{pages}{481} (\bibinfo{year}{2005}).

\bibitem[{\citenamefont{Greiner et~al.}(2005)\citenamefont{Greiner, Regal,
  Stewart, and Jin}}]{greiner_regal_05}
\bibinfo{author}{\bibfnamefont{M.}~\bibnamefont{Greiner}},
  \bibinfo{author}{\bibfnamefont{C.~A.} \bibnamefont{Regal}},
  \bibinfo{author}{\bibfnamefont{J.~T.} \bibnamefont{Stewart}},
  \bibnamefont{and} \bibinfo{author}{\bibfnamefont{D.~S.} \bibnamefont{Jin}},
  \bibinfo{journal}{Phys. Rev. Lett.} \textbf{\bibinfo{volume}{94}},
  \bibinfo{pages}{110401} (\bibinfo{year}{2005}).

\bibitem[{\citenamefont{Holstein and Primakoff}(1940)}]{holstein_primakoff_40}
\bibinfo{author}{\bibfnamefont{T.}~\bibnamefont{Holstein}} \bibnamefont{and}
  \bibinfo{author}{\bibfnamefont{H.}~\bibnamefont{Primakoff}},
  \bibinfo{journal}{Phys. Rev.} \textbf{\bibinfo{volume}{58}},
  \bibinfo{pages}{1098} (\bibinfo{year}{1940}).

\bibitem[{\citenamefont{Jordan and Wigner}(1928)}]{jordan_28}
\bibinfo{author}{\bibfnamefont{P.}~\bibnamefont{Jordan}} \bibnamefont{and}
  \bibinfo{author}{\bibfnamefont{E.}~\bibnamefont{Wigner}},
  \bibinfo{journal}{Z. Phys.} \textbf{\bibinfo{volume}{47}},
  \bibinfo{pages}{631} (\bibinfo{year}{1928}).

\bibitem[{\citenamefont{Rey et~al.}(2006{\natexlab{a}})\citenamefont{Rey,
  Satija, and Clark}}]{rey_satija_06a}
\bibinfo{author}{\bibfnamefont{A.~M.} \bibnamefont{Rey}},
  \bibinfo{author}{\bibfnamefont{I.~I.} \bibnamefont{Satija}},
  \bibnamefont{and} \bibinfo{author}{\bibfnamefont{C.~W.} \bibnamefont{Clark}},
  \bibinfo{journal}{J. Phys. B}
  \textbf{\bibinfo{volume}{39}}, \bibinfo{pages}{S177}
  (\bibinfo{year}{2006}{\natexlab{a}}).

\bibitem[{\citenamefont{Rey et~al.}(2006{\natexlab{b}})\citenamefont{Rey,
  Satija, and Clark}}]{rey_satija_06b}
\bibinfo{author}{\bibfnamefont{A.~M.} \bibnamefont{Rey}},
  \bibinfo{author}{\bibfnamefont{I.~I.} \bibnamefont{Satija}},
  \bibnamefont{and} \bibinfo{author}{\bibfnamefont{C.~W.} \bibnamefont{Clark}},
  \bibinfo{journal}{Phys. Rev. A} \textbf{\bibinfo{volume}{73}},
  \bibinfo{pages}{063610} (\bibinfo{year}{2006}{\natexlab{b}}).

\bibitem[{\citenamefont{Rey et~al.}(2006{\natexlab{c}})\citenamefont{Rey,
  Satija, and Clark}}]{rey_satija_06c}
\bibinfo{author}{\bibfnamefont{A.~M.} \bibnamefont{Rey}},
  \bibinfo{author}{\bibfnamefont{I.~I.} \bibnamefont{Satija}},
  \bibnamefont{and} \bibinfo{author}{\bibfnamefont{C.~W.} \bibnamefont{Clark}},
  \bibinfo{journal}{New J. Phys.} \textbf{\bibinfo{volume}{8}},
  \bibinfo{pages}{155} (\bibinfo{year}{2006}{\natexlab{c}}).

\bibitem[{\citenamefont{Rousseau et~al.}(2006)\citenamefont{Rousseau, Arovas,
  Rigol, H\'ebert, Batrouni, and Scalettar}}]{rousseau_arovas_06}
\bibinfo{author}{\bibfnamefont{V.~G.} \bibnamefont{Rousseau}},
  \bibinfo{author}{\bibfnamefont{D.~P.} \bibnamefont{Arovas}},
  \bibinfo{author}{\bibfnamefont{M.}~\bibnamefont{Rigol}},
  \bibinfo{author}{\bibfnamefont{F.}~\bibnamefont{H\'ebert}},
  \bibinfo{author}{\bibfnamefont{G.~G.} \bibnamefont{Batrouni}},
  \bibnamefont{and} \bibinfo{author}{\bibfnamefont{R.~T.}
  \bibnamefont{Scalettar}}, \bibinfo{journal}{Phys. Rev. B}
  \textbf{\bibinfo{volume}{73}}, \bibinfo{pages}{174516}
  (\bibinfo{year}{2006}).

\bibitem[{\citenamefont{Rigol et~al.}(2006)\citenamefont{Rigol, Muramatsu, and
  Olshanii}}]{rigol_muramatsu_06}
\bibinfo{author}{\bibfnamefont{M.}~\bibnamefont{Rigol}},
  \bibinfo{author}{\bibfnamefont{A.}~\bibnamefont{Muramatsu}},
  \bibnamefont{and} \bibinfo{author}{\bibfnamefont{M.}~\bibnamefont{Olshanii}},
  \bibinfo{journal}{Phys. Rev. A} \textbf{\bibinfo{volume}{74}},
  \bibinfo{pages}{053616} (\bibinfo{year}{2006}).

\bibitem[{\citenamefont{Horstmann et~al.}(2007)\citenamefont{Horstmann, Cirac,
  and Roscilde}}]{horstmann_cirac_07}
\bibinfo{author}{\bibfnamefont{B.}~\bibnamefont{Horstmann}},
  \bibinfo{author}{\bibfnamefont{J.~I.} \bibnamefont{Cirac}}, \bibnamefont{and}
  \bibinfo{author}{\bibfnamefont{T.}~\bibnamefont{Roscilde}},
  \bibinfo{journal}{Phys. Rev. A} \textbf{\bibinfo{volume}{76}},
  \bibinfo{pages}{043625} (\bibinfo{year}{2007}).

\bibitem[{\citenamefont{Mathey et~al.}(2009)\citenamefont{Mathey, Vishwanath, ,
  and Altman}}]{mathey09}
\bibinfo{author}{\bibfnamefont{L.}~\bibnamefont{Mathey}},
  \bibinfo{author}{\bibfnamefont{A.}~\bibnamefont{Vishwanath}},
  \bibnamefont{and} \bibinfo{author}{\bibfnamefont{E.}~\bibnamefont{Altman}},
  \bibinfo{journal}{Phys. Rev. A} \textbf{\bibinfo{volume}{79}},
  \bibinfo{pages}{013609} (\bibinfo{year}{2009}).

\end{thebibliography}
\end{document}